\documentclass[global,twocolumn]{svjour_arxiV}

\usepackage[numbers,square,sort&compress]{natbib}
\usepackage[latin1]{inputenc}
\usepackage[dvips]{graphicx}
\usepackage{graphicx}
\usepackage{amsmath,amsbsy,amsfonts,amssymb}
\usepackage{bm}
\usepackage{epsfig}

\begin{document}

\title{A high resolution time-of-flight mass spectrometer for the detection of
       ultracold molecules}

\author{Stephan D. Kraft\inst{1}\thanks{s.kraft@fzd.de, now at Forschungszentrum Dresden-Rossendorf, Germany} \and
Jochen Mikosch\inst{1} \and Peter Staanum\inst{1,2}\thanks{now at
Dept. of Physics and Astronomy, University of Aarhus, Denmark} \and
Johannes Deiglmayr\inst{1}  \and J\"org Lange\inst{1} \and Andrea
Fioretti\inst{3} \and Roland Wester\inst{1} \and Matthias
Weidem\"uller\inst{1}\thanks{m.weidemueller@physik.uni-freiburg.de}}

\institute{Physikalisches Institut, Universit\"at Freiburg,
Hermann-Herder-Stra{\ss}e 3, 79104 Freiburg, Germany  \and Institut
f\"ur Quantenoptik, Universit\"at Hannover, Welfengarten 1, 30167
Hannover, Germany \and Istituto per i Processi Chimico-Fisici,
C.N.R., via Moruzzi 1, 56124 Pisa, Italy }

\date{\today}

\maketitle

\begin{abstract}

We have realized a high-resolution time-of-flight mass spectrometer combined
with a magneto-optical trap. The spectrometer enables excellent optical access
to the trapped atomic cloud using specifically devised acceleration and
deflection electrodes. The ions are extracted along a laser beam axis and
deflected onto an off axis detector. The setup is applied to detect atoms and
molecules photoassociated from ultracold atoms. The detection is based on
resonance-enhanced multi-photon ionization. Mass resolution up to $m/\Delta
m_{\rm rms} = 1000$ at the mass of $^{133}$Cs is achieved. The performance of
this spectrometer is demonstrated in the detection of photoassociated
ultracold $^7$Li$^{133}$Cs molecules near a large signal of $^{133}$Cs ions.

\end{abstract}

\section{Introduction}\label{sec:intro}

Applying ultracold atoms as targets for collision experiments with
electrons, atoms, ions or photons requires sensitive ion
detectors~\cite{Ullrich}. As an important example, in the study of
ultracold ground state molecules formed from ultracold atoms an
efficient and state-selective molecule detection scheme is based on
resonance-enhanced multi-photon ionization
(REMPI)~\cite{Petty-REMPI} in combination with time-of-flight mass
spectrometry.

Unlike many other experiments involving mass spectrometers, experiments with
ultracold gases require a very good optical access.  For example, in our setup
we use more than 15 laser beams with beam diameters of up to 1.5 cm coming
from different directions for trapping and manipulating the atomic cloud. Mass
spectrometers employed to date in experiments with ultracold gases use simple
electric field geometries for the extraction and detection of the ions. They
offer good optical access at the expense of poor mass resolution ($m/\Delta m
\simeq 10$), which nevertheless is sufficient to separate homonuclear alkali
dimers~\cite{Fioretti-Csmolecules, Gabbanini-Rbmolecules, Nikolov-Kmolecules}
and in some cases also heteronuclear dimers from their atomic
constituents~\cite{Kerman-RbCs,Mancini-KRbmolecules,Wang-KRb,Bigelow-NaCs}.
For photoassociation experiments of heteronuclear molecules with an unbalanced
mass ratio of the atomic constituents or for the separation of different
isotopomers, e.g. $^{39}$K$^{85}$Rb and $^{39}$K$^{87}$Rb, a high mass
resolution is required. In order to discriminate nearby masses with largely
different signal amplitudes this mass resolution needs to be near 10$^3$.

Here we present an advanced time-of-flight mass spectrometer with higher
resolution and good optical access for the detection of ions formed by
ionization of trapped, ultracold atoms or molecules. High mass resolution is
achieved by time focusing of the ions following a Wiley-McLaren
scheme~\cite{Wiley-Mclaren}. The field plates for accelerating the ions and
the detector allow for optimal optical access to the region of the
magneto-optical trap (MOT) as described in more detail below. Our motivation
for this design was the selective detection of ultracold $^7$Li$^{133}$Cs
molecules on a large background of $^{133}$Cs atoms, which we recently
achieved with this spectrometer \cite{KraftLiCs}.

This article is organized as follows. In Sec.~\ref{sec:theory} the
design of the time-of-flight spectrometer is presented. The
experimental characterization is described in
Sec.~\ref{sec:characterization}. In Sec.~\ref{sec:lics} we
demonstrate the detection of ultracold LiCs molecules. A discussion
is given in Sec.~\ref{sec:discussion}.

\section{Design of the mass spectrometer}\label{sec:theory}

A time-of-flight (TOF) mass spectrometer~\cite{Weickhardt-Review}
offers several advantages in experiments with cold atoms and
molecules. With this type of spectrometer it is possible to record
the whole mass spectrum at once, which is not the case for
quadrupole or magnetic mass filters~\cite{Paul-Spektrometer}. This
allows one to monitor the development of the ratio of atoms to
molecules, which can be helpful in optimizing production mechanisms
for molecules. In addition, it is insensitive to magnetic fields,
which are needed to operate magneto-optical or magnetic traps. As
demonstrated by Wiley and McLaren~\cite{Wiley-Mclaren} and employed
in numerous experiments, e.g., with gas cells or molecular beams,
the resolution of a time-of-flight mass spectrometer can be
significantly improved by using a two step acceleration of the ions
in electric fields of different strength created between three
parallel field plates. By adjusting the ratio between the two field
strengths it is possible to employ a time-focus, i.e. to minimize
the arrival time spread of an ion cloud of given mass on an ion
detector.

\begin{figure}[h]
\includegraphics[width=0.8\linewidth]{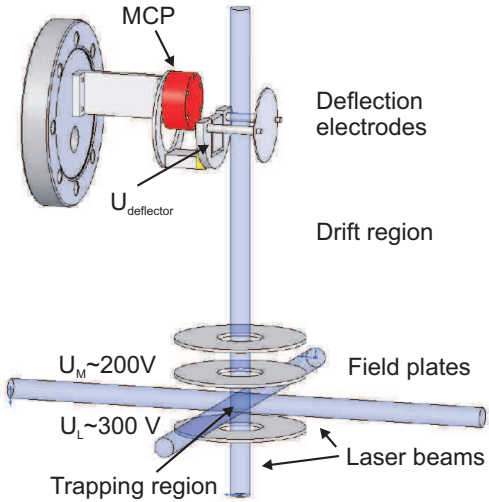}
\caption{\label{fig:layout} Drawing of the mass spectrometer (for dimensions
see text). The ions are produced from a magneto-optical trap between the lower
two plates and accelerated upwards. After drifting through a field free region
of 30 cm they are deflected onto a microchannel plate detector. The setup
allows optical access in the whole horizontal plane and along the vertical
axis. The laser beams for the magneto-optical trap are indicated by three
cylinders.  More than 10 additional laser beams (not shown) also intersect the
trapping region.}
\end{figure}

As shown in Fig.~\ref{fig:layout} the spectrometer for cold gases
consists of two parts: an acceleration and a detection region. The
acceleration is done by three horizontal field plates built around
the magneto-optical trap from which the ions are produced. The
stainless steel field plates are 2\,mm thick, have a diameter of
80\,mm, with a central hole of 30\,mm diameter. No grids are mounted
across the holes. The lower and the middle plate are separated by
40\,mm and centered vertically around the MOT. The upper electrode
is separated from the middle one by 20\,mm. A voltage
$U_L=300\,\rm{V}$ is applied to the lower plate and voltages around
$U_M=200\,\rm{V}$ to the middle plate. The upper plate is grounded.
This accelerates the ions upwards through the holes into the drift
region.

After passing the field free drift region of 30\,cm length, the
ions arrive at the detection part of the spectrometer. This part
is composed of a deflection electrode and a microchannel plate
(MCP) detector attached on a single CF63 flange. The detector is
mounted off the vertical axis to allow for optical access along
this axis. The voltage applied to the deflector is typically
-1.2\,kV. A grounded grid is placed in front of the MCP to avoid
background gas ions being pulled onto the detector.

As ion detector we use a matched pair of MCP plates (Burle
Industries)~\cite{Wiza-MCP}. The front of the first plate is held at a
potential of -2.1\,kV, the back of the second plate at \mbox{-100\,V}. The
produced electrons hit an anode plate behind the MCP connected to ground
through a 100\,k$\rm\Omega$ resistor. AC signals from the anode are coupled
out through a capacitance of 1\,nF and referenced with 50\,$\rm\Omega$ to
ground. A single ion arriving at the MCP produces a pulse of about 7\,ns width
at an amplitude of about 3\,mV.  The deflection electrode is mounted
electrically isolated from the grounded plates using PEEK, which is a high
quality plastic and sustains bake-out temperatures over 200$^\circ$C without
mechanical distortion and outgassing.

This design leaves full optical access in the entire horizontal
plane, in planes tilted up to 30$^\circ$ from the horizontal and
along the vertical axis. Compared to the theoretical resolution of
$m/\Delta m \simeq 4 \cdot 10^4$ given in~\cite{Wiley-Mclaren} in
our design the resolution is limited to about $10^3$ by two effects
as shown by simulations with SIMION~\cite{SimIon}. The main decrease
in resolution of about one order of magnitude is due to the
deflection of the ions onto the detector in the upper part of the
spectrometer. In addition, the configuration of the field plates
around the trapping region leads to a factor of three decrease in
mass resolution, because the large distance between the field plates
and the large holes to accommodate the laser create a rather
inhomogeneous acceleration field. To avoid this effect one could
think of mounting grids across the holes. This would lead to
interference effects of the laser beams passing through which would
make an efficient cooling of the atoms impossible. Enlarging the
outer diameter of the plates would reduce the optical access to the
MOT which is needed for more than 15 laser beams including a 140\,W
CO$_2$ laser beam passing through the atomic cloud in the horizontal
plane.

A more sophisticated  electrostatic mirror such as a
reflectron~\cite{Boesl-Reflectron} would eliminate the influence of
the deflector in the upper part of the spectrometer. In our
experiment a free path of about 4 cm diameter in the vertical axis
is necessary for 5 laser beams with 0.8 cm waist radius each. This
makes the design of a grid-less
reflectron~\cite{Frey-Reflectron,Schmid-Reflectron} very difficult
and not worthwhile in an experiment with ultracold atoms since the
already achieved resolution is more than sufficient.

\section{Characterization}\label{sec:characterization}

\begin{figure}[h]
\includegraphics[width=\linewidth]{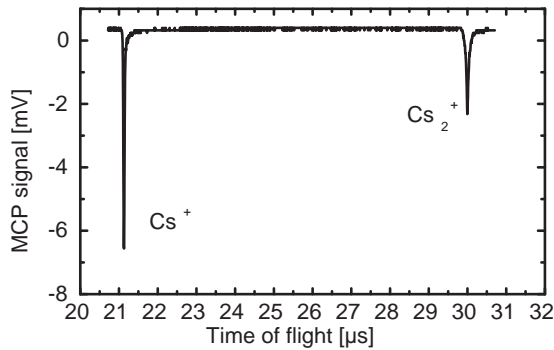}
\caption{\label{fig:TOF} A measured TOF spectrum of Cs$^+$ and
Cs$_2^+$. The atomic ions arrive at the detector at 21.2\,$\mu$s,
the molecular ions at 29.7\,$\mu$s.}
\end{figure}

For the characterization of the mass spectrometer we produced
$\rm{Cs}^+$ ions from atoms in a MOT by photoionization with a
pulsed dye laser (Radiant Dyes, Narrowscan, 7\,ns pulse,
20\,mJ/pulse, 20\,Hz, operated at 716\,nm). The MOT is loaded from a
Zeeman-slowed atomic beam from a Cs oven. For these measurements,
typically $10^7$\,Cs atoms are trapped at a particle density of
$10^{10}\,\rm{cm}^{-3}$. In addition $\rm{Cs}_2$ molecules are
formed via photoassociation of the atoms using a cw Ti:Sapphire
laser (Coherent MBR 110) and ionized through REMPI using the pulsed
dye laser~\cite{Fioretti-Csmolecules}. Fig.~\ref{fig:TOF} shows a
typical measured time-of-flight spectrum for Cs$^+$ and Cs$_2^+$.
The atomic ions arrive 21.2\,$\mu$s after the ionization laser
pulse, the flight time of the molecular ions is about 1.4  times
longer, as expected from the square root of the mass ratio.

\begin{figure}[h]
\includegraphics[width=\linewidth]{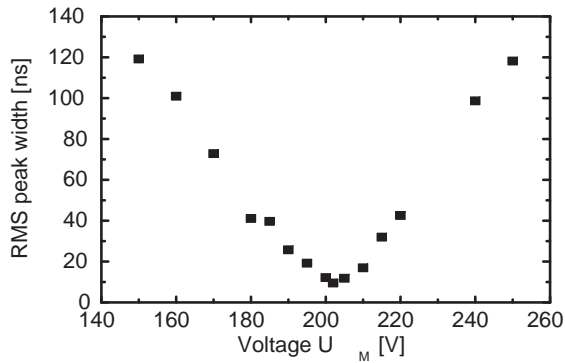}
\caption{\label{fig:focus} Time focusing of the ions on the detector
by varying the voltage of the second field plate. A minimum width of
10\,ns is achieved at a voltage of 202\,V.}
\end{figure}

To optimize the time focusing of the ions on the detector, we
vary the ratio of the electric field strengths by applying a fixed
voltage of $U_L$ = +300\,V on the lowest field plate, keeping the
upper one on ground and varying the voltage $U_M$ applied to the
middle plate. This measurement was done with only very few ions
produced per pulse ($<$ 15), to avoid interaction among the ions
through Coulomb repulsion and saturation effects of the MCP.
Fig.~\ref{fig:focus} shows the rms peak width of the atomic ion
signal for the atomic ions as a function of $U_M$. The width is
extracted from an average of 128 single TOF traces. The signal
depends strongly on the voltage $U_M$ with a minimal rms peak
width of about 10\,ns at $U_M =$ 202\,V. The peak width
corresponds to a mass resolution of $\frac{m}{\Delta m_{\rm rms}} =
1000$. This is in good agreement with the width of 15\,ns obtained
from the SIMION simulations. Fig. \ref{fig:focus} shows that the
voltages applied to the field plates have to be controlled on the
one percent level for having optimal resolution.

Broadening of the signal due to Coulomb repulsion among the ions
on their way to the detector is negligible for small ion numbers.
However, it becomes the dominant broadening process for larger ion
numbers. The influence of the Coulomb repulsion is investigated in
two distinct ways. First, the number of ions produced in the
extraction field is varied in a controlled way, second, a constant
number of ions is produced in a field free environment and a
variable delay between production and extraction of these ions is
applied.

\begin{figure}[h]
\includegraphics[width=\linewidth]{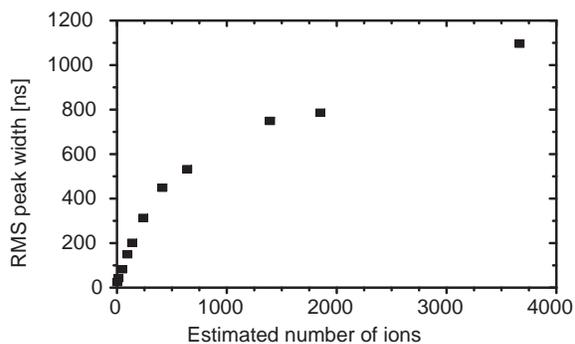}
\caption{\label{fig:resolution}Peak width of the MCP signal for
different ion numbers.}
\end{figure}

The number of ions, which is estimated from the peak area of the
MCP signal, is varied by changing the power of the dye laser. The
peak width increases with increasing ion number as shown in
Fig.~\ref{fig:resolution}, hence the mass resolution decreases.
The larger the ion number, the larger is the space charge density
which leads to a stronger repulsion in the ionic cloud. Therefore
the peak width on the detector grows with the number of produced
ions. Since the ion number is varied by two orders of magnitude,
special care must be taken to prevent saturation of the MCP for
large ion numbers. Hence, we lowered the voltage applied to the
MCP for larger ion intensities and used a relative calibration in
order to compare the MCP signals at different MCP voltages.

\begin{figure}[h]
\includegraphics[width=\linewidth]{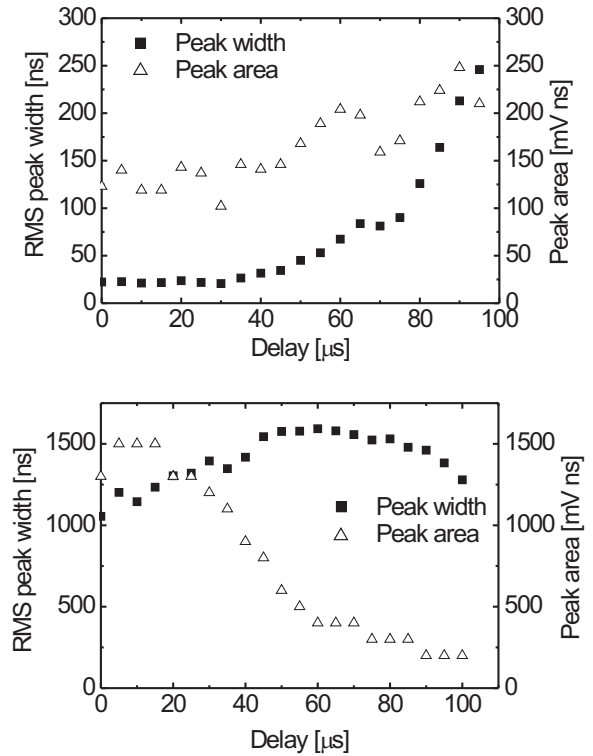}
\caption{\label{fig:coulomb} Coulomb repulsion for about 10 ions
(upper) and more than 150 ions (lower).}
\end{figure}

In a second measurement at a constant ion number, a variable delay is inserted
between the time of production of the ions (time of dye laser pulse) and the
beginning of their acceleration towards the detector. The voltages on the
lower and middle field plates are switched with fast HV-switches (Behlke
Electronic) and delays of 0-100\,$\mu$s are generated with a delay generator
(Berkeley Nucleonics Corporation, Model 555 Pulse/Delay Generator). This gives
the ion cloud more time to expand under the influence of the Coulomb force and
leads to a broader temporal distribution of the ions on the detector. Since
the temperature of the ions is only a few hundred $\mu$K thermal energy can be
neglected and the expansion is due to the Coulomb interaction only. For small
ion numbers almost no effect is visible for small delay times (upper graph in
Fig.~\ref{fig:coulomb}). For delay times beyond 40\,$\mu$s the signal becomes
broader starting from less than 25\,ns up to almost 250\,ns after 95\,$\mu$s
delay. One would expect the peak area to be constant for a constant ion
number. The observed increase is due to a reduced saturation of the MCP with
increased delay time. This is here already the case for low ion numbers
because of the small phase space volume occupied by the ions created from
ultracold atoms. For large delays the ions hit a bigger area of the MCP and
hence saturation is reduced. This saturation effect implies that the small rms
peak widths measured at small delay times may even be overestimated. For
large ion numbers (lower graph in Fig.~\ref{fig:coulomb}) the peak width
at zero delay is much larger than for the small ion numbers because the ions
repel each other significantly on their way to the detector. The insertion of
a delay has only little influence on the peak width, however, the total number
of detected ions decreases rapidly. Under these conditions the strong Coulomb
repulsion enlarges the cloud too much to map all ions onto the detector any
longer and some of them hit the electrodes or the chamber walls.

\section{Detection of ultracold LiCs molecules}\label{sec:lics}

In this section the mass-spectrometric detection of ultracold LiCs molecules,
recently observed for the first time \cite{KraftLiCs}, is described in detail.
Due to the small LiCs signal levels, we switch for this purpose to a single
ion counting configuration of the detector. Ultracold LiCs molecules are
produced via photoassociation in a two species MOT containing Li and Cs
atoms. Photoassociation is achieved by tuning a cw Ti:Sapphire laser ($1/e^2$
radius of 0.24\,mm, power\,=\,680\,mW) to molecular resonances identified in
previous spectroscopic studies~\cite{staanum}, which leads to large signals in
contrast to Ref. \cite{KraftLiCs} where MOT light induced the
photoassociation. The molecules are then ionized by two photons at 696\,nm
from a dye laser.

\begin{figure}[h]
    \begin{center}
        \includegraphics[width=0.82\linewidth]{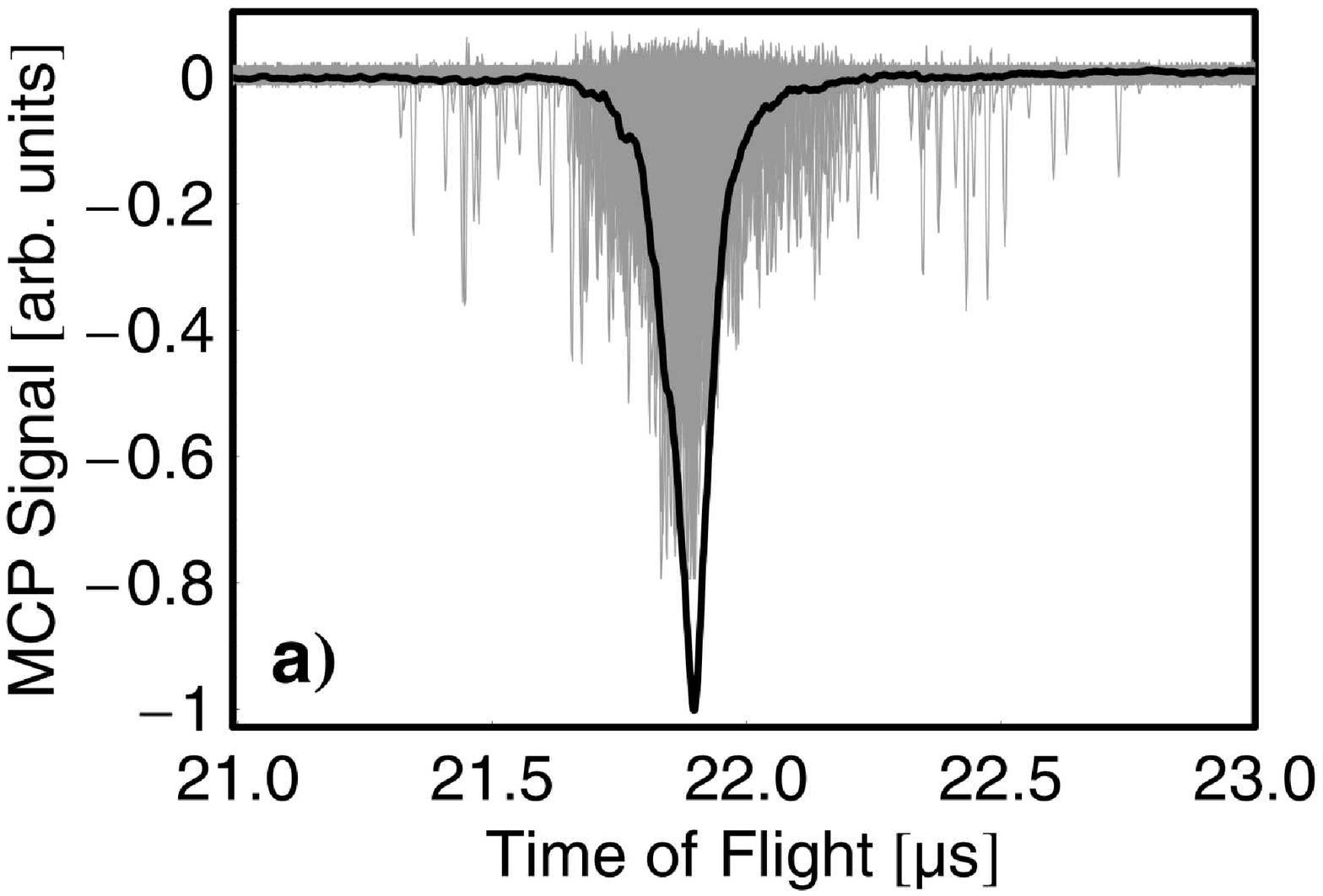}
        \includegraphics[width=0.82\linewidth]{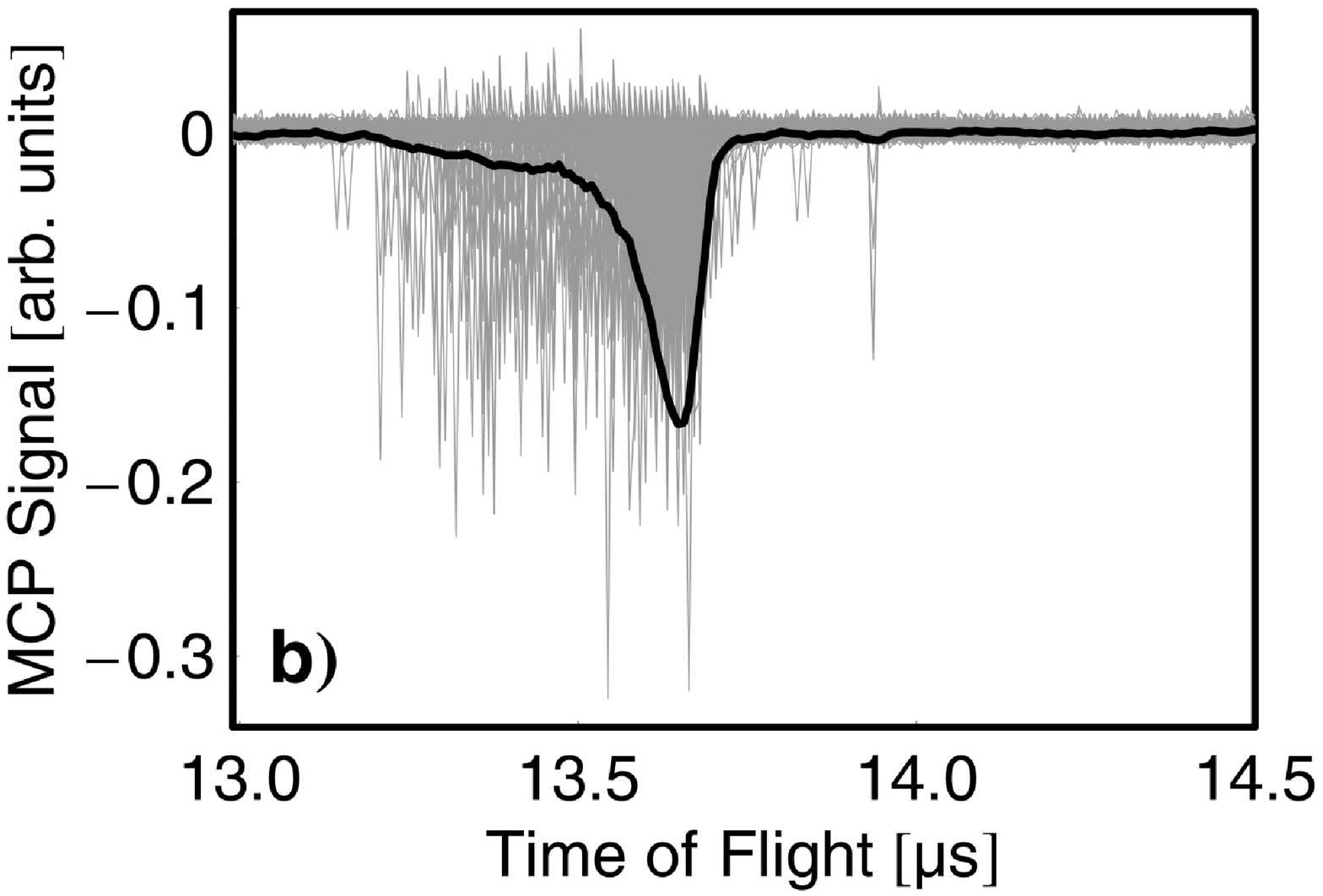}
    \end{center}
\caption{\label{fig:cs_fastions} Time of flight trace around the arrival time
of Cs$^+$ ions. Shown are an overlay of 350 individual traces (grey) and the
averaged signal (black) (enlarged by a factor of 10 for visibility). Two
different set of voltages are applied for a) optimal time focusing and b)
increase selectivity for masses above $^{133}$Cs}
\end{figure}

In order to reduce the number of Cs$^+$ background ions arriving at the
detector, we block the Cs-MOT lasers 2\,ms before ionization using mechanical
shutters. This guarantees that all Cs atoms are in the 6S$_{1/2}$ ground state
and therefore three photons are necessary for their
ionization. Fig.\,\ref{fig:cs_fastions}a shows 350 single TOF traces of a pure
Cs$^+$ sample plotted on top of each other together with the average
signal. In this plot two different features are visible: a sharp peak at
21.90$\mu$s and a broad distribution of ions arriving roughly from 21.3$\mu$s
to 22.8$\mu$s.  While the sharp peak (width $\sim 80$ns) is caused by
initially cold Cs atoms from the MOT, we attribute the broad distribution
(width $\sim 500$ns) to fast Cs ions produced by ionization of
photo-dissociated Cs$_2$ molecules\,\footnote{These Cs$_2$ molecules are
formed from cold Cs atoms by photoassociation through the cooling
lasers.}~\cite{KraftLiCs}. A Cs$_2$ molecule can be dissociated by a photon of
the ionization laser transforming it into a pair of Cs atoms either in the
6S$_{1/2}$+6P$_{3/2}$ or the 6S$_{1/2}$+6P$_{1/2}$ states. Each cesium atom
moves with 0.18\,eV or 0.22\,eV kinetic energy, depending on the
asymptote. Ionization of the fast Cs(6p) atoms by two additional photons leads
to ions arriving up to about 300\,ns earlier or later compared to the
time-of-flight peak of the cold cesium atoms. The distribution of the arrival
times reflects the angular distribution of the fragment velocities of the
Cs(6p) atoms that emerge from the Cs$_2$ photodissociation. The measurement of
such arrival time distributions is expected to open up new opportunities for
the study of quantum-state specific properties of ultracold molecules.

These fast ions constitute a significant background at the expected
arrival time of 22.47$\,\mu$s for the LiCs ions. We therefore
modified the voltage settings in the following way: larger absolute
values of voltages lead to an improved energy focusing; changing the
ratio of the field plate voltages $U_L$/$U_M$ allows us to shift the
arrival of the fast ions to earlier times at the cost of
none-optimal time focusing. Fig.\,\ref{fig:cs_fastions}b shows a
modification ($U_L$=800\,V, $U_M$=610\,V) where the arrival time of
nearly the entire signal produced by fast Cs$^+$ ion has been
shifted to a time before the cold Cs$^+$ ions arrive. Therefore it
is now possible to detect even small numbers of ions with a slightly
higher mass than Cs$^+$ with almost no background contribution from
the fast Cs$^+$ ions.

\begin{figure}[h]
\begin{center}
\includegraphics[width=0.82\linewidth]{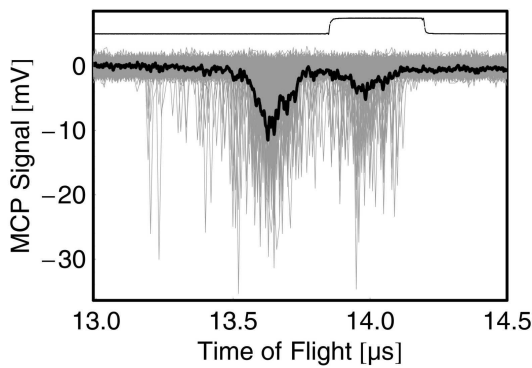}
\end{center} \caption{\label{fig:lics_tof} Overlay of 600 individual time
of flight traces (grey), averaged signal (black) and window set for detection
of LiCs$^+$ (dashed line). }
\end{figure}

Fig.\,\ref{fig:lics_tof} shows the two well separated signals from roughly
1,000 Cs$^+$ and about 300 LiCs$^+$ ions. For the counting of LiCs molecules,
only the ions appearing in a gated time window of 350\,ns around the expected
arrival time of LiCs$^+$ are detected by a fast discriminator and a
counter. The gate was chosen carefully to avoid a background caused by fast
Cs$^+$ ions. The modified set of voltages reduces the number of fast Cs$^+$
ions arriving in the time window set for detection to less than one per 100
shot which allows us to measure even small numbers of formed LiCs molecules.

\section{Discussion}\label{sec:discussion}

The mass spectrometer for ultracold atomic and molecular gases presented here
yields optimal optical access to the trap region with a mass resolution of
$\frac{m}{\Delta m} = 1000$ in good agreement with design simulations. For
large ion numbers, however, the mass resolution is decreased due to Coulomb
repulsion between the ions during their flight time. Therefore, one has to
adjust the ion intensity to optimize signal strength versus mass resolution
and suppress unwanted ions (e.g., atomic cesium). In this work this has been
achieved by blocking the Cs excitation lasers prior to ionization. A further
reduction can either be done by pushing away the atoms with a resonant laser
before ionizing the molecules or by switching on the deflection electrode
after the atomic ions have passed through the deflector region.

We demonstrate the capability of this setup to distinguish between
atomic Cs$^+$ and molecular LiCs$^+$ ions produced from a
two-species MOT ~\cite{Schloeder} setup via photoassociation. This
spectrometer will be useful for experiments studying ultracold
chemical reactions, such as the quantum-state resolved detection of
products of the Li~+~Cs$_2$ exchange reaction.

This work is supported by the Deutsche Forschungsgemeinschaft in the
Schwerpunktsprogramm 1116 "Interactions in Ultracold Atomic and Molecular
Gases" under WE-2661/1-2 and 1-3. P. S. and J. D. are supported by the EU
Research Training Network "Cold Molecules" under the Contract No
HPRN-2002-00290. The stay of A. F. in Freiburg was supported by the CATS
network of the European Science Foundation and the Cold Molecules
Network. J. D. acknowledges financial support from the German-French
University, Saarbr\"ucken.

\end{document}